\documentclass[12pt,preprint]{aastex}

\shorttitle{SN 2006aj and low-luminosity GRBs}
\shortauthors{Cobb et al.}

\begin{document}

\title{SN 2006aj and the nature of low-luminosity gamma-ray bursts 
}

\author{B.~E. Cobb\altaffilmark{1}, C.~D. Bailyn\altaffilmark{1}, P.~G. van Dokkum\altaffilmark{1}, and
P. Natarajan\altaffilmark{1}}
\email{cobb@astro.yale.edu}

\altaffiltext{1}{Department of Astronomy, Yale University, P.O. Box 208101, New Haven, CT 06520}

\begin{abstract}
We present SMARTS consortium optical/IR light curves of SN 2006aj associated with GRB 060218.  We find that this event is
broadly similar to two previously observed events SN 1998bw/GRB 980425 and SN 2003lw/GRB 031203.  In
particular, all of these events are greatly under-luminous in gamma-rays compared
to typical long-duration GRBs.  We find that the observation by \textit{Swift}
of even one such event implies a large enough true event rate to create difficulties
in interpreting these events as typical GRBs observed off-axis.  Thus these
events appear to be intrinsically different from and much more common than high-luminosity 
GRBs, which have been observed in large numbers out to a redshift
of at least 6.3. 
The existence of a range of intrinsic energies of GRBs may present challenges
to using GRBs as standard candles. 
\end{abstract}


\section{Introduction}
While some long-duration gamma-ray bursts (GRBs) are clearly associated with supernovae (SNe),
a deeper understanding of the GRB/SN connection remains elusive.  The GRB/SN link
was first confirmed observationally with the detection of the
low-reshift GRB 980425/SN 1998bw ($z=0.0085$) \citep{Galama+98}.  
GRB 980425, however, was not a typical GRB; it was
under-luminous in gamma-rays and had no detected optical afterglow (OAG).
SNe were later associated with typical GRBs at cosmological redshifts \citep[e.g.][]{Bloom+99,DellaValle+03}.
However, another low gamma-ray luminosity event similar to GRB 980425/SN 1998bw was not observed until GRB 031203.  
This burst was also orders of magnitude under-energetic and, despite its low redshift ($z=0.1055$) \citep{Prochaska+04},
was followed by only a dim OAG \citep{Malesani+04}.
Follow-up observations of this burst detected a SN-like brightening \citep{Cobb+04,Gal-Yam+04,Thomsen+04}
and SN 2003lw was confirmed spectroscopically by \cite{Tagliaferri+04}.  
The spectra of SN 2003lw were reminiscent of those of SN 1998bw \citep{Malesani+04}.  The
The two SN also had similiar peak magnitudes, although SN 2003lw was somewhat brighter 
and evolved more slowly.  Their light curve shapes were also qualitatively
different, with SN 1998bw climbing smoothly to peak while SN 2003lw experienced
a broad plateau.

The low gamma-ray luminosity of GRB 980425 and 031203 suggested that they might represent a new GRB category.
Alternatively, they could be normal GRBs that appear under-luminous because their 
jetted emission is observed off-axis \citep[e.g.][]{YYN03,Ramirez+05}.
A comparison of the event rate of low- to typical-luminosity GRBs was warranted 
but only a small and inhomogeneous sample of
well-localized GRBs existed before \textit{Swift}, as pre-\textit{Swift} GRBs were detected
using multiple instruments, each with unique sensitivity and sky coverage.  
\textit{Swift} provides a large and homogeneous GRB sample that is well-suited for rate calculations.

On 2006 February 18 at 03:34:30 UT \textit{Swift} detected a new low-luminosity event: GRB 060218 \citep{Cusumano+06}.
This was an unusual GRB with weak gamma-ray emission lasting over 2000 seconds
\citep{Barthelmy+06}. This burst was followed by an unusual OAG
that brightened for 10 hours before decaying like a typical OAG \citep[e.g.][]{Marshall+06}.
This was the first \textit{Swift} GRB to be associated with a SN: SN 2006aj.
The SN was initially noted in spectral observations \citep{Masetti+06}
and then detected as an optical re-brightening \citep[e.g.][etc.]{D'Avanzo+06,Ovaldsen+06}.
At $z=0.033$, GRB 060218 is now the second closest GRB with a measured redshift \citep{MH06}.
This is the third example of a GRB-related SN
in which the gamma-rays are highly under-luminous and
the SN light curve is clearly distinct from the GRB's OAG.
Hereafter, we will refer to these Long-duration, Low-Luminosity events as L$_3$--GRBs.

SMARTS observations of SN 2006aj began on 2006 Feb 22 at 00:35 UT \citep{CB06}.  We present optical/IR 
data obtained between 5 and 30 days following GRB 060218. 
Our homogeneous data demonstrate that the light curve of SN 2006aj is
qualitatively similar to that of the pre-\textit{Swift} L$_3$-GRB SNe 1998bw/2003lw. 
We argue in \S\,4 that the detection of this
single event in the \textit{Swift} era already places strong constraints on
the nature of L$_3$-GRBs.

\section{Observations and Data Reduction}
Our data was obtained using the ANDICAM 
instrument mounted on the 1.3m telescope at Cerro Tololo Inter-American 
Observatory.\footnote{http://www.astronomy.ohio-state.edu/ANDICAM}
This telescope is operated as part of the Small and Moderate Aperture Research
Telescope System (SMARTS) consortium.\footnote{http://www.astro.yale.edu/smarts}
Nightly imaging was obtained over 26 days with occasional interruptions
for weather and equipment problems.
The GRB/SN was only observable for a limited period of time immediately after twilight ($\lesssim1$hr).
Consequently, all observations were obtained at high airmass ($\sec{(z)}\gtrsim 2$).

Each nightly data set consisted of 6 individual 360-second I-band observations 
obtained simultaneously with 30 dithered 60-second J-band images.  The data were reduced 
in the same way as in \cite{Cobb+04}.  A few additional steps were added, including 
cosmic ray removal in the I-band images using the L.A.
Cosmic program\footnote{http://www.astro.yale.edu/dokkum/lacosmic/} \citep{vanDokkum01} 
and I-band fringe correction using an iterative masking technique.
Some images were not included in the final frames because of 
excessive background due to twilight or because of telescope
drift.  Typically,  the
final frames were equivalent to 30 minutes of I/J-band
exposure time.

The relative magnitude of the SN + host galaxy was determined by comparison with 11 (3) on-chip, non-variable
objects in I (J) using seeing matched aperture photometry.  Differential magnitudes were converted to apparent
magnitudes by comparison, on photometric nights, with Landolt
standard stars in the fields of RU149 and PG1047 \citep{Landolt92} for the I-band
images, and with 3 on-chip 2MASS stars \citep{Skrutskie+06} for the J-band images.
The difference in airmass value between the Landolt standard frames and science frames
was corrected for using an extinction coefficient of 0.066 magnitudes per airmass.

The light curves are shown in Figure 1 and the photometric data are summarized in Table 1.
The error bars represent the photometric measurement error, which accurately reflects nightly
variations in image quality but does not account for systematic measurement errors.
In addition to the relative night-to-night uncertainty, there is a systematic
error of 0.05 magnitudes in I and J resulting from uncertainties in 
the photometric calibration.

\section{Results}
Figure 1 shows that GRB 060218's optical and IR counterpart brightened for the first two weeks and then 
proceeded to gradually decay.  This behavior is not consistent with that of a standard GRB OAG \citep[e.g.][]{Tagliaferri+05} 
but is reminiscent of the low redshift events SNe 1998bw/2003lw.  Identification of this
optical emission as a SN is possible from our data alone due to our dense observations and
the object's particular transient behavior; spectral evidence obtained by
other groups clearly supports our claim \citep{Modjaz+06,Sollerman+06,Mirabal+06}. 
The well-sampled nature of our observations
allows us to determine an unambiguous time of peak brightness in I and J.  
This parameter will be important for determining the
amount of mass ejected in the supernova explosion, thought this modeling
is beyond the scope of this paper.
The position of peak brightness was determined by fitting second order cubic 
splines to the data points, with errors derived from the formal chi-squared error on 
the fits in combination with the error on the measured magnitudes.
The combination of the host galaxy and
SN reaches a peak apparent magnitude in I of $16.91\pm0.05$ mag after $13.1_{-1.9}^{+2.1}$ days  
and in J of $16.65\pm0.06$ mag after $17.6_{-3.2}^{+3.5}$ days. 
The rest-frame time to peak is, therefore, approximately 12.7 days in I and 17.0 days in J.

The Galactic extinction correction along the line of sight to the host galaxy
is taken to be $A_I=0.23$ mag and $A_J=0.11$ mag, assuming the Galactic extinction curves of \cite{CCM89} 
and a measured reddening value of E(B-V)=0.127 mag \citep{Guenther+06}.
The pre-burst SDSS model magnitude of 
the host galaxy is $i=19.805\pm0.041$ mag, not corrected for Galactic extinction \citep{Adelman+06}.
Using the transformation equations derived by \cite{Lupton05}, this corresponds to
$I=19.368\pm0.047$ mag.
The peak absolute magnitude of the supernova is, therefore, $M_I=-19.02\pm0.09$ mag.
No k-correction has been applied, but this should result in minimal error due
to the low redshift of the burst.  A correction of -0.04 magnitudes was applied to account for
spectral stretching.  
The exact pre-burst J magnitude of the host is unknown as the host galaxy is too dim
to appear in the 2MASS catalog.  Our observations indicate the host galaxy must have a J magnitude
$>18$. Assuming a range of host magnitudes from
18 mag to 20 mag, the peak absolute magnitude of the supernova in J is approximately
$M_J =-19.1\pm0.2$ mag. 

Note that 2006aj clearly
peaks later in J than in I.  This later peak at redder wavelengths
follows the trend seen in SN 1998bw, which, in the rest-frame, peaked 1.6 days earlier in V than in I.
The rest-frame V-band peak of SN 2006aj occurred at approximately 9.7 days \citep{Modjaz+06}, 
which is 2.9 days prior to the I-band peak.  Likewise,
SN 2003lw peaked in V at ~$\sim18$ days \citep{Malesani+04} and in I at $\sim23$ days \citep{Cobb+04,Malesani+04}.
The combined light of the galaxy and the SN reddens from I-J = 0.0 mag during the first week to I-J = 0.6 mag
for the last few observations.  
This is a stronger evolution in I-J color than experienced by
either SN 1998bw or SN 2003lw, whose I-J colors in the
first month only change by about 0.3 magnitudes \citep{Gal-Yam+04}.  This comparison
is complicated, however, by the unknown intrinsic I-J color of the host galaxy
of SN 2006aj. 

\section{Discussion}
Our data, together with those of \cite{Modjaz+06}, \cite{Sollerman+06} and \cite{Mirabal+06}, 
show that GRB 060218 is the third detected GRB to be followed by a dominant SN.  
With such a small sample being used to extrapolate the characteristics
for an entire category, it is important to collect
homogeneous and detailed observations over a wide range of wavelengths. 
Our data provides dense IR coverage, which
extends the total SN 2006aj dataset out to redder bands than reported thus far.
It is instructive to compare all three cases in
which L$_3$--GRBs have been detected, each
of which was followed by a Type Ic SN: 1998bw, 2003lw and 2006aj (see Figure 2).
We note that the limit on observing L$_3$--GRB events are more 
stringent than those on observing the associated SNe,
so it is unlikely that GRBs for which no optical 
counterpart are observed are of this character.
The properties of these three events are shown in Table 2.
All three SNe are very similar in peak brightness, though 2003lw may
be half a magnitude brighter than the others.
The biggest difference
between the bursts is their rise times, with 2006aj peaking the fastest
and 2003lw taking the longest time to peak.
The rest-frame photon energy at which the GRB spectrum peaks ($E_{p,i}$) appears to increase
with increasing SN rise time.

As discussed in the introduction, it would be instructive to compare
the frequency of these L$_3$ bursts with that of high-luminosity bursts
using the homogeneous \textit{Swift} dataset.  Such a comparison
is now possible since GRB 060218 is the first \textit{Swift}-detected burst that falls
in the category of low-luminosity GRBs associated with Type Ic SNe.
   The low observed fluxes and redshifts of L$_3$--GRBs suggest
   that the underlying rate of L$_3$--GRB events may be quite high \citep[see also][]{Pian+06,Soderberg+04}, since the
   volume in which these sources can be observed is much smaller 
   than that of typical GRBs.  The ratio of the event rate of
   L$_3$--GRBs to ordinary long-duration GRBs is expected to be 
\[\frac{R_{int,~L_3}}{R_{int,~GRB}} = \frac{R_{obs,~L_3}}{R_{obs,~GRB}} \times \left(\frac{D_{c,~GRB}}{D_{c,~L_3}}\right)^3\]
   where $R_{int}$ denotes the true rate per comoving volume of the
   two kinds of events, $R_{obs}$ is the observed event rate seen by
   a given experiment, and $D_c$ is the comoving radial distance out to which the events could
   be observed by that experiment.  In the case of \textit{Swift}, the 
   limiting volume in which L$_3$--GRBs are observable is so small that the
   observation of even one of these events implies an extremely high
   ratio of true rates.

   Specifically, GRB 060218 would not have triggered the BAT event
   monitor if it had been $\sim2$ times fainter, which corresponds to a 
   maximum redshift of $z=0.046$.  By contrast, the 31 high-luminosity, long-duration \textit{Swift} GRBs 
 that have measured redshifts have an average redshift of $z=2.6$. 
   Many of these bursts would have been detected even
   if they had occurred at $z\gtrsim6$.
   We will take a conservative approach, however, and assume that most of these bursts would not have been
   detected had they been at extreme redshifts and adopt $z=2.6$
   as the redshift below which the high-luminosity GRB samples are complete.
   Assuming the concordance cosmology
   of $\Omega_\Lambda = 0.73$, $\Omega_M = 0.27$ and a Hubble constant of
   71 km s$^{-1}$ Mpc$^{-1}$, the ratio of the comoving radial distance cubed for these two events is $3.1 \times 10^4$.  
   Assuming that the
   ratio of rates is equal to the number of long-duration GRB events observed to date
   (1 L$_3$--GRB per about 100 GRBs), we find a true event ratio of $3\times10^2$.
In the near future, thanks to \textit{Swift}, the cosmological GRB
sample may be complete out to at least a redshift of $z\sim5$.  For this volume, 
the true event ratio increases by a factor of $\sim2$.  
If the current highest GRB redshift measurement, $z=6.3$, is representative of
the redshift out to which we have a complete sample, then the
true event ratio is $\sim10^3$.  We also note that we are assuming a constant
GRB rate per unit time.  If, instead, the rate scales with the star
formation rate \citep[e.g.][]{Natarajan+05}, then the relevant ratio would be higher still.

   These event ratios provide a constraint on the
   origins of L$_3$--GRBs.  One explanation of these
   events is that they are standard GRBs observed off-axis
   \citep[e.g.][]{YYN03,Ramirez+05}.  This is an attractive option as it
   accounts for all long-duration GRBs using a ``unified model'',
   with observed differences attributed only to viewing angle.
   However, this scenario implies a maximum true
   rate ratio, which would be generated if the L$_3$--GRBs could
   be observed from any angle.  This upper limit is
   $2\pi / \theta_j^2$, where $\theta_j^2$ is the jet opening solid angle of
   a typical GRB (in steradians).  Jet breaks observed in
   X-ray/optical AG light curves constrain the jet opening angle to $\sim 10^\circ$ \citep[e.g.][]{Frail+01},
   so we infer a maximum true rate ratio of 65 in this model.  This ratio
   is a factor of five lower than the ratio of $3\times10^2$, which
   we inferred for $z=2.6$, and lower by $\sim20$ than
   for $z=6.3$.  

Since \textit{Swift} has observed only one L$_3$--GRB,
the underlying event rate remains uncertain; GRB 060128/SN 2006aj could
have been a serendipitous event. From Poisson statistics, the 90\% lower confidence limit of
the true event rate is an order of magnitude lower
than assumed above, which implies that the off-axis scenario is still acceptable,
provided one assumes a limit of $z=2.6$ and a constant GRB rate.
If \textit{Swift} detected a second L$_3$--GRB, the off-axis scenario would
become significantly less probable.  Such a Poissonian analysis addresses
the question ``given an event rate, what is the probability of seeing
an event'', whereas in this case one might more appropriately
ask the Bayesian question ``having seen an event, what
is the probability of a given event rate''.  Assuming 
a uniform prior on the distribution of event rates (an assumption for which there
is no real basis), we find that the off-axis scenario is implausible
at the 98\% level. Of course, the undetermined luminosity function of GRBs could
complicate this calculation, as could cosmic evolution of the GRB
source population.

At face value, however, the above calculation of the event rate
suggests that there is a category of GRB events that is intrinsically
different from that of typical GRBs.   
   Several suggestions have been made
   for how these differences can be accounted for including
   the possibility that the gamma-rays are produced in
   supernova shock breakout \citep{MM99,TMM01}
   or ``failed collapsars'' in which highly relativistic
   jets fail to develop due to baryon loading \citep{WM99}.
The orientation-corrected energies of GRBs have been claimed
to be constant at $\sim10^{51}$ ergs \citep{Frail+01}. However,
if intrinsically low-energy GRBs exist as a separate
population, efforts
to use GRBs as standard candles \cite[e.g.][]{Lazzati+06,Ghirlanda+04} may be compromised. 

\acknowledgments
We thank SMARTS observers D. Gonzalez and J. Espinoza for their dedication
and S. Tourtellotte for assistance with
optical data reduction. We greatly appreciate discussions
with A. Cantrell and J. Emerson.  This work is supported by NSF
Graduate Fellowship DGE0202738 to BEC and NSF/AST grant 0407063 and \textit{Swift}
grant NNG05GM63G to CDB.

\clearpage
\begin{deluxetable}{rrrr}
\tablecolumns{3}
\tablewidth{0pc}
\tablecaption{Photometry of SN 2006aj in the Host Galaxy of GRB 060218}
\tablehead{
\colhead{Days after GRB\tablenotemark{a}} & \colhead{I magnitude\tablenotemark{b}}   & \colhead{J magnitude\tablenotemark{b}}}
\startdata
          4.88 & $17.51\pm0.01$ & $17.26\pm0.03$ \\
          5.87 & $17.34\pm0.01$ & $17.20\pm0.03$ \\
          6.87 & $17.22\pm0.01$ & $17.08\pm0.03$ \\
\enddata
\tablenotetext{a}{Days after burst trigger at 2006 Feb 18, 03:34:30 UT.}
\tablenotetext{b}{These values have not been corrected for Galactic extinction.
There is an additional uncertainty of 0.05 mag in the
transformation of relative to apparent magnitudes.}
\tablecomments{The complete version of this table is in the electronic edition of
the Journal.  The printed edition contains only a sample.}
\end{deluxetable}

\begin{deluxetable}{lcccc}
\tablecolumns{4}
\tablewidth{0pc}
\tablecaption{GRB/SN Properties}
\tablehead{\colhead{}                                   &
           \multicolumn{1}{c}{GRB 980425}               &
           \multicolumn{1}{c}{GRB 031203}               &
           \multicolumn{1}{c}{GRB 060218} 	        &
           \colhead{}                                   \\
           \colhead{}                                   &
           \colhead{SN 1998bw}                          &
           \colhead{SN 2003lw}                          &
           \colhead{SN 2006aj}                          }
\startdata
	redshift 			  		     & 0.0085		 	        & 0.1055				& 0.033 			\\
	fluence ($10^{-6}$ erg cm$^{-1}$) 		     & $2.8\pm0.5$\tablenotemark{a}     & $2.0\pm0.4$\tablenotemark{b} 		& $6.8\pm0.4$\tablenotemark{c} 	\\
	total duration (s) 		 		     & $\sim40$    		        & $\sim40$ 				& $>2000$ 			\\ 
	$E_{p,i}$ (keV) 		  		     & $55\pm15$\tablenotemark{d}       &   $158\pm51$\tablenotemark{d} 	& $<10$\tablenotemark{e}	\\
	$E_{iso}$ ($1\times10^{50}$ erg)\tablenotemark{f}    & $0.010\pm0.002$\tablenotemark{d} & $1.0\pm0.4$\tablenotemark{d}		& $0.65\pm0.15$\tablenotemark{e}\\
        I-band T$_{peak}$ (days)\tablenotemark{g} 	     & $17.7\pm0.3$\tablenotemark{h}    & $18 - 28$\tablenotemark{i} 		& $12.7_{-1.8}^{+2.0}$	\\
        peak $M_I$ (mag)				     & $-19.27\pm0.05$\tablenotemark{h} & $-19.0$ to $-19.7$\tablenotemark{i} 	& $-19.02\pm0.09$		\\
     I-J SN color, $\sim$T$_{peak,~I}$ 			     & 0.5\tablenotemark{l}             & $\sim0.4$\tablenotemark{i} 		& $\sim0.0$\tablenotemark{m}				\\
\enddata
\tablenotetext{a}{(40 - 700 keV) \cite{Pian+00}}
\tablenotetext{b}{(20 - 200 keV) \cite{SLS04}}
\tablenotetext{c}{(15 - 150 keV) \cite{Sakamoto+06}}
\tablenotetext{d}{\cite{Amati06}}
\tablenotetext{e}{\cite{Amati+06}}
\tablenotetext{f}{rest-frame 1 - 10,000 keV, assuming $H_0=70$ km s$^{-1}$ Mpc$^{-1}$, $\Omega_\Lambda=0.7$, $\Omega_M=0.3$}
\tablenotetext{g}{in the rest-frame}
\tablenotetext{h}{\cite{Galama+98}}
\tablenotetext{i}{exact value depends strongly on extinction assumptions, \cite{Cobb+04,Gal-Yam+04,Malesani+04,Thomsen+04}}
\tablenotetext{l}{at $\sim22$ days post-burst, \cite{Patat+01}}
\tablenotetext{m}{assuming a J-band host magnitude of 19}
\end{deluxetable}

\begin{figure}
\includegraphics[width=1\textwidth]{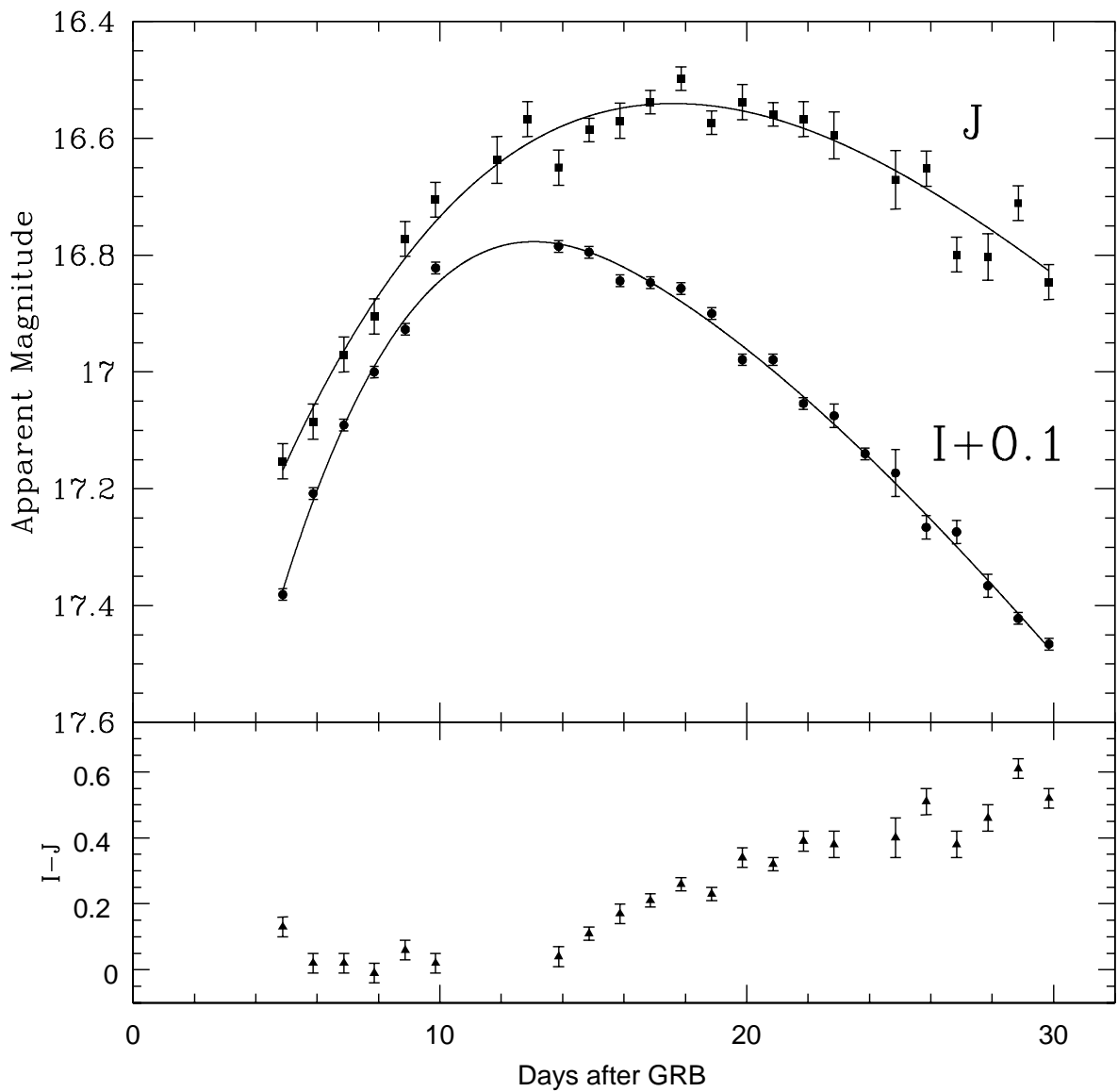}
\caption{\textit{Top panel:} The I-band (\textit{circles}) and J-band (\textit{squares})
aperture photometry light curve of the host+SN of GRB 060218.  Values have been
corrected for a Galactic foreground extinction of $A_V$=0.39 mag. For clarity, 
the I-band points have been shifted by +0.1 magnitudes.  Error bars are photometric
measurement errors and do not include possible systematic effects.
The curves are fit with second order cubic splines.
\textit{Bottom panel:} I-J color evolution, the combined light is observed to redden with time.}
\end{figure}

\begin{figure}
\includegraphics[width=1\textwidth]{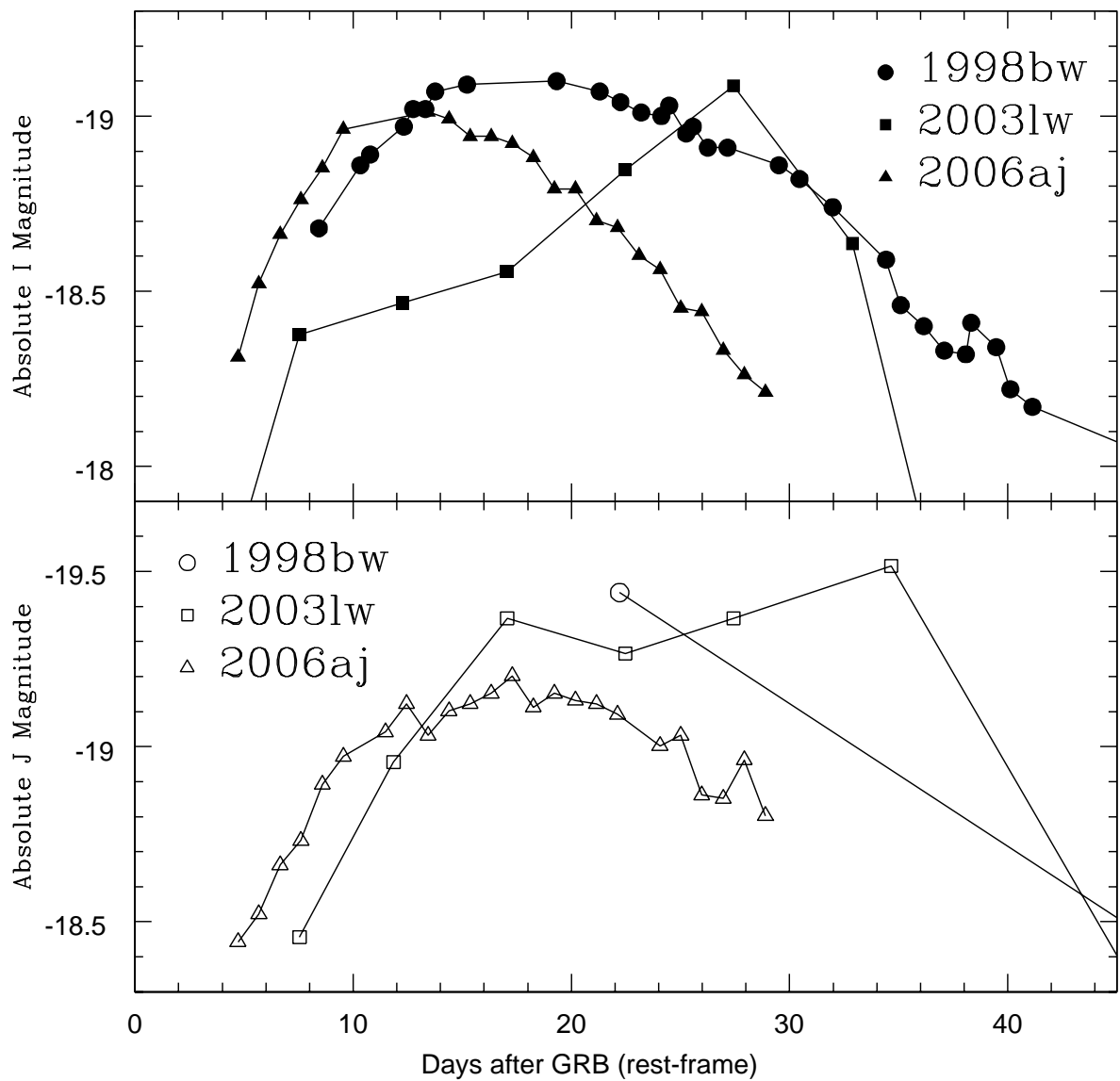}
\caption{Absolute magnitude light curves of SN 1998bw (\textit{circles}),
SN 2003lw (\textit{sqaures}) and SN 2006aj (\textit{triangles}) in
I (\textit{top panel}) and J (\textit{bottom panel}).  The SN 2003lw
data has been binned in intervals of 5 days.
The apparent J-band host
galaxy magnitude of SN 2006aj is assumed to be 19 mag.  SN 2003lw
may be shifted by $\sim-0.5$ magnitudes if a stronger line-of-sight
extinction is assumed.}
\end{figure}

\end{document}